\documentclass[preprint,proceedings]{rmaa}

\usepackage{paralist}

\usepackage{psfig,color}

\SetYear{2003}
\SetConfTitle{Compact Binaries in the Galaxy and Beyond}

\title{A New View of the Supersoft X-Ray Source Cal 87 Observed with XMM-Newton} 

\author{
  M. Orio,\altaffilmark{1,2} 
  K. Ebisawa,\altaffilmark{3}
  J. Heise,\altaffil{4}
  and W. Hartmann\altaffilmark{4}}

\altaffiltext{1}{INAF -- Turin Astronomical Observatory, Italy.}
\altaffiltext{2}{Department of Astronomy,  U Wisconsin at Madison, USA.}
\altaffiltext{3}{INTEGRAL Science Data Centre, Versoix,  Switzerland.}
\altaffiltext{4}{SRON National Institute for Space Research, Utrecht,
 The Netherlands.}

\shortauthor{Orio, Ebisawa, \& Heise}
\shorttitle{Cal 87 observed with XMM-Newton}

\fulladdresses{
\item Ken Ebisawa:
INTEGRAL Science Data Centre, Chemin d'Ecogia, 16 CH-1290 Versoix, Switzerland
  (\email{ebisawa@obs.unige.ch}).
\item John Heise and Wouter Hartmann:SRON National Institute for Space Research, Sorbonnelaan 2,
3584 CA Utrecht, Netherlands.
(\email{jheise@purple.sron.nl,W.Hartmann@sron.nl.}).
\item Marina Orio: INAF -- Turin Astronomical Observatory,
              Strada Osservatorio 20, I-10025 Pino Torinese (TO), Italy
         and
Department of Astronomy, 475 N. Charter Str., Madison WI
              53706, USA.
(\email{orio@astro.wisc.edu}) 
}

\listofauthors{W. J. Henney, A. Collaborator, \& L. Author}
\indexauthor{Orio, M.}
\indexauthor{Ebisawa, K.}
\indexauthor{Heise, J.}
\indexauthor{Hartmann, J.}

\abstract{Cal 87 was observed with with {\sl XMM-Newton}
in April of 2003. The source shows a rich emission spectrum, 
where lines can be identified if they are red-shifted by 700-1200
km s$^{-1}$. These lines seem to have been emitted in a wind
from the system. The eclipse is observed 
to be shifted in phase  by 0.03 $\phi_{\rm orb}$, where  $\phi_{\rm orb}$
is the phase of the optical light curve.}

\addkeyword{Stars: binaries, white dwarfs}
\addkeyword{X-rays: stars}

\begin{document}
\maketitle

\section{The observed X-ray spectrum}

Cal 87 was observed with {\sl XMM Newton} on  April 18-19 2003,
 for 21.8 hours (two full orbital cycles). 
The background corrected count rate measured in the two
{\sl RGS-1} and {\sl RGS-2}
instruments in the 0.33-2.5 \AA \ range is 0.0764$\pm$0.0012
and 0.0653$\pm$0.0011 cts s$^{-1}$, respectively.
 The spectrum clearly appears to be an emission line one.
There is an obvious  similarity with the X-ray grating spectra
of this source,
(Motch et al. 2002, Bearda et al. 2002),
 however Cal 87 displays a much softer spectrum.
Tentative line identification for the emission
features in the {\sl RGS} spectra indicates a red shift
of several identifiable  lines with velocities in the 700-1200
km s$^{-1}$ range. We conclude that these lines must be originated in a wind.
The {\sl EPIC}
spectra, despite pile-up effects that cannot be completely
corrected for, offer a broader energy range (0.2-10 keV),
and allow us to conclude that the central white dwarf is 
not visible at any time, even outside of eclipse.
\begin{figure}[!t]
  \includegraphics[angle=-90,width=\columnwidth]{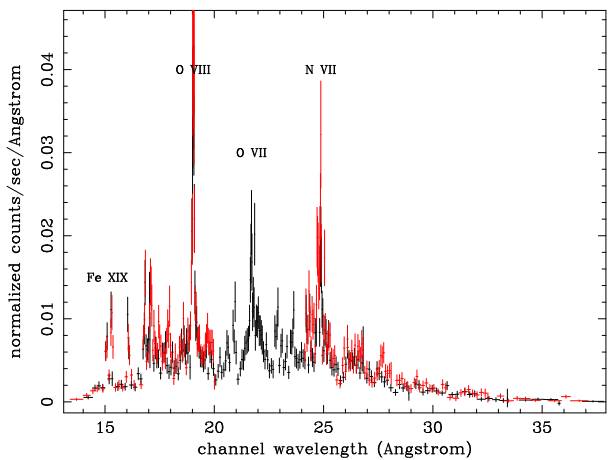}
  \caption{The spectrum observed with  the two {\sl RGS} gratings. The
{\sl RGS-1} spectrum is shown in black, the {\sl RGS-2} spectrum is in red.
Some line identifications are labelled. 
No significant signal above the background is detected at wavelengths below
15 \AA. }
  \label{fig:simple}
\end{figure}
\section{The light curve}

The eclipse already observed in X-rays is observed again with 
{\sl XMM-Newton}. It
is more definite and deeper in the {\sl EPIC} in the
{\sl RGS} light curve,
not only because of the better S/N, but
also because the depth
of the eclipse is greater at lower energy.
 It is shifted by $\Delta \phi_{\rm orb}$=0.03 with respect
to the eclipse observed at optical wavelengths. 
We speculate that this may be so 
because the X-rays are emitted by the Accretion Disk Corona, while 
the optical radiation originates instead from the disk.


\begin{thebibliography}

\bibitem{} Bearda,  H., et al. 2002, A\&A, 385, 511

\bibitem{} Motch, C., Bearda H., and Neiner C. 2002, 385, 91
 
 
\end{thebibliography}
\end{document}